\documentclass{article}

\usepackage{arxiv}

\usepackage{url}            
\usepackage[ruled,noline]{algorithm2e}
\usepackage{amsmath}
\usepackage{amssymb}
\usepackage{amsfonts}
\usepackage{caption}
\usepackage[colorinlistoftodos]{todonotes}
\usepackage{lineno}
\usepackage{mathtools}
\usepackage{float}
\usepackage{url}            

\usepackage{etoolbox}
\providetoggle{fullBackground}

\usepackage{parskip}  

\newcommand{\xy}{\hspace{-0.5pt}x\hspace{-0.75pt}y}

\DeclareFontFamily{U}{wncy}{}
\DeclareFontShape{U}{wncy}{m}{n}{<->wncyr10}{}
\DeclareSymbolFont{mcy}{U}{wncy}{m}{n}
\DeclareMathSymbol{\comb}{\mathord}{mcy}{"58} 

\usepackage[T1]{fontenc}
\DeclareFontFamily{T1}{calligra}{}
\DeclareFontShape{T1}{calligra}{m}{n}{<->s*[1.44]callig15}{}
\DeclareMathAlphabet\mathcalligra   {T1}{calligra} {m} {n}

\usepackage{accents}
\newlength{\dhatheight}

\title{multiMap: A Gradient Spoiled Sequence for Simultaneously Measuring $B_1^+$, $B_0$, $T_1/M_0$, $T_2$, $T_2^\ast$, and Fat Fraction of a Slice}

\author{
  Nicholas Dwork\thanks{www.nicholasdwork.com, nicholas.dwork@ucsf.edu} \\
  Department of Radiology and Biomedical Imaging \\
  University of California in San Francisco
    \And
  Adam B. Kerr \\
  Center for Cognitive and Neurobiological Imaging \\
  Stanford University
    \And
  Ethan M. I. Johnson \\
  Department of Biomedical Engineering \\
  Northwestern University
    \And
  Corey A. Baron \\
  Robarts Research Institute \\
  The University of Western Ontario
    \And
  Shreyas S. Vasanawala \\
  Department of Electrical Engineering \\
  Stanford University
    \And
  Peder E. Z. Larson \\
  Department of Radiology and Biomedical Imaging \\
  University of California in San Francisco
    \And
  Adam M. Bush \\
  Department of Electrical Engineering \\
  Stanford University
    \And
  John M. Pauly \\
  Department of Electrical Engineering \\
  Stanford University
}

\begin{document}
\maketitle

\begin{abstract}
  We propose multiMap, a single scan that can generate several quantitative maps simultaneously.
  The sequence acquires multiple images in a time-efficient manner, which can be modeled for $T_2$, $T_2^\ast$, main- and transmit-field inhomogeneity, $T_1$:equilibrium magnetization, and water and fat content.  The sequence is constructed so that cross-dependencies between parameters are isolated.  Thus, each parameter can be estimated independently.
  Estimates of all parameters are shown on bottle phantoms, the brain, and the knee.
  The results are compared to estimates from established techniques.
\end{abstract}


\section{Introduction}
Several standard scans can be employed to measure quantities of interest in patients: double angle mapping for measuring $B_1$; inversion recovery for measuring $T_1$; spin echo imaging for measuring $T_2$; and an IDEAL sequence for measuring off-resonance, fat fraction, and $T_2^\ast$ \cite{reeder2004multicoil,reeder2007water,lu2008multiresolution}.  However, these scans are time consuming and delay any treatment of the patient's condition. Significant research has been conducted in quantifying these parameters quickly.  For example, the variable flip angle method of $T_1$ mapping is much faster than using inversion recovery.  Others have attempted to estimate several quantities from a single scan; for example, in \cite{metere2017simultaneous}, Metere et al. developed a single scan to quantify $T_1$, $T_2^\ast$, and susceptibility.

Alternatively, the MR fingerprinting sequence can be used for quantification.  Fingerprinting also attempts to quantify several parameters simultaneously \cite{ma2013magnetic,bipin2018magnetic}.  A sequence of random parameters (including flip angles, times between flip angles, times before readout, and the trajectories used during acquisition) is used to generate a catalog of expected signals and to image.  For each voxel imaged, the measured signal is compared to the catalog of simulated signals; the parameters that generated the closest signal are assigned to the voxel.

While fingerprinting shows great promise, there remain several obstacles to overcome.  The size of the signal catalog grows exponentially in the number of parameters to quantify.  This makes storage of the dictionary difficult, and can lead to very long reconstruction times.  The problem of storage is somewhat addressed with compression of the catalog, but remains an issue.  Furthermore, compression exacerbates the reconstruction problem (because the algorithm for reconstruction becomes more computationally expensive).  Additionally, fingerprinting is not a steady state sequence, so the entire volume of interest must be imaged with each undersampled acquisition.  This leads to significant artifacts which one hopes are incoherent enough that they can be overcome during the quantification process.

The Saturated Double Angle Method (SDAM) is a method of performing $B_1$ double angle mapping without waiting for recovery between readout acquisitions (thus reducing the scan time) \cite{cunningham2006saturated}.  A non-selective $B_1$-insensitive adiabatic saturation pulse is followed by some time for recovery before an imaging radio-frequency (RF) pulse is emitted.  The sequence is repeated for two different RF pulses ($60^\circ$ and $120^\circ$) from which the Double Angle Method of $B_1$ mapping can be performed \cite{insko1993mapping}.  Due to the saturation pulse, the $M_z$ component is (in general) not fully recovered at the time of the imaging RF pulse, so the signal-to-noise ratio (SNR) of the quantification is reduced.

We realized that SDAM could be augmented to quantify many more parameters.  In this work, we present multiMap: a gradient spoiled sequence that combines several of the standard methods of quantification into a single scan.  multiMap quantifies $B_1$, $B_0$, $T_1/M_0$, $T_2$, $T_2^\ast$, and fat-fraction.  The main insight that led to multiMap is that altering the phase after saturation in SDAM does not affect the recovery state at the time of the second RF pulse.  Thus, the transverse magnetization that results from the saturation pulse, which is normally ignored, can be used for additional quantification.  Furthermore, the transverse state after the SDAM imaging RF pulse can also be used for further interrogation.  We detail the theory and methods of this technique in the sections that follow.

\section{Methods}

\subsection{Background}
\label{sec:background}

In this subsection, we discuss the mathematical models that we employed for quantification with multiMap.

The magnetic vector associated with a sample $M=(M_x,M_y,M_z)\in\mathbb{R}^3$ is governed by Bloch's equations \cite{nishimura1996principles}.  Let $M_{\xy}\in\mathbb{C}$ such that $M_{\xy}=M_x+i M_y$ where $i=\sqrt{-1}$.
After excitation, assuming the sample contains water and fat, $M_{\xy}$ behaves according to \cite{reeder2007water}
\begin{equation}
  M_{\xy}(t) = \exp\left( i\Delta\omega_0 t \right) \left[
    \exp\left( -t/T_{2,W}^\ast \right) W + \exp\left( i\omega_{cs} t - t/T_{2,F}^\ast \right)  F
    \right],
  \label{eq:mxyModel}
\end{equation}
where $t$ is the time since excitation (assuming negligible transverse magnetization prior to excitation),
$\Delta\omega_0\in\mathbb{R}$ is the off-resonance frequency due to differences in the main magnetic field, $T_{2,W}^\ast> 0$ is the decay constant of the water in the sample, $T_{2,F}^\ast> 0$ is that of fat, and $\omega_{cs}\in\mathbb{R}$ is the chemical shift frequency of fat (which may have a non-zero phase at time $t=0$).  The values $W\in\mathbb{C}$ and $F\in\mathbb{C}$ represent the water and fat in the sample.

If $180^\circ$ spin pulses are used after excitation, then off-resonance frequencies (due to differences in the main magnetic field or chemical shift) are compensated prior to data acquisition.  This leads to a different exponential rate of decay $T_2>T_2^\ast$ and is modelled by $M_{\xy}(t) = M_{\xy}(0) \exp\left(-t/T_2\right)$.  After excitation, the $M_z$ component of the magnetization vector recovers according to $M_z(t) = M_z(t_0) + M_0\left( 1 - \exp\left( -(t-t_0)/T_1 \right) \right)$, where $t_0$ is the time of excitation, $t>t_0$, $T_1$ is the recovery constant, and $M_0$ is the proton density of the sample.

Selective excitation is used to excite a slice of a volume \cite{nishimura1996principles}.  When doing so, the signal at each two-dimensional location $(x,y)$ in the slice equals
\begin{equation}
  M_{\xy}(x,y,t) = \int_{-\infty}^{\infty} M_{\xy}(x,y,z,t) \, dz.
  \label{eq:sliceIntegral}
\end{equation}
We are left to determine the signal intensity as a function of $z$ location in the slice for each pixel's location.  The equations presented above that dictate behavior of $M$ are constant across the slice.  To determine the effect of the RF pulses across the slice, the RF pulse is approximated as a piecewise constant function \cite{pauly1991parameter}.  Then, the rotation matrix for each piece as a function of $z$ can be determined analytically.  The magnetic vector at location $z$ in a voxel after excitation is related to the magnetic vector prior to excitation according to 
\begin{equation*}
  \begin{bmatrix}
    M_x(z) \\ M_y(z) \\ M_z(z)
  \end{bmatrix}^+ = R_M(z) \, R_{M-1}(z) \, \cdots R_2(z) \, R_1(z) \begin{bmatrix}
    M_x(z) \\ M_y(z) \\ M_z(z)
  \end{bmatrix} = R(z) \begin{bmatrix}
    M_x(z) \\ M_y(z) \\ M_z(z)
  \end{bmatrix}
\end{equation*}
where $R_j$ is the rotation matrix of the $j^{\text{th}}$ non-zero piece of the RF pulse, and $R=R_M R_{M-1}\cdots R_2 R_1$ is the composite rotation matrix.

One can use the above equations to model the behavior of the RF pulses, relaxation, and recovery as a function of $z$ position in the slice.  Then, one can estimate the value of $M_{\xy}$ at some location $(x,y)$ by approximating \eqref{eq:sliceIntegral} with a Riemann sum calculated over an interval $[z_{\min},z_{\max}]$ centered on the slice.
We will utilize this technique several times throughout this paper.



\subsection{multiMap Sequence and Quantitation}
\label{sec:methods}

\begin{figure}[ht]
  \centering
  \includegraphics[width=0.9\linewidth]{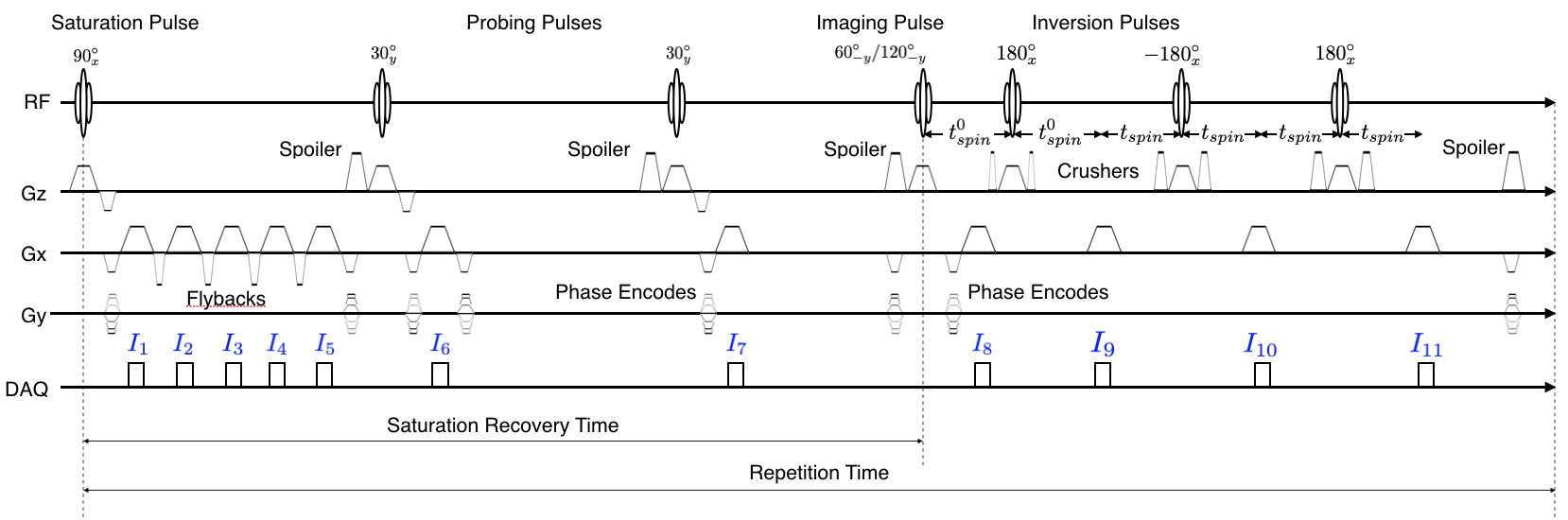}
  \protect\caption{The pulse sequence diagram for the multiMap sequence.  For each line acquired, the repetition is repeated twice: once with an imaging pulse of $60^\circ$ and once with $120^\circ$.}
  \label{fig:multiMapSeq}
\end{figure}

The complete multiMap sequence is shown in Fig. \ref{fig:multiMapSeq}.  A Cartesian spin-warp (2DFT) trajectory is used for acquisition.  A repetition of the sequence consists of seven RF pulses (which are labeled with flip angles in Fig. \ref{fig:multiMapSeq}, named the saturation pulse, the probing pulses, the imaging pulse, and the inversion pulses) and eleven acquisitions.  The index of the image that corresponds to each acquisition is labeled in blue above the acquisition waveforms in Fig. \ref{fig:multiMapSeq}.  For each line of the trajectory, there are two repetitions (two segments): one for an imaging pulse with a flip angle of $60^\circ$ and another for an imaging pulse with a flip angle of $120^\circ$.

The second inversion pulse is negated to account for inaccuracy in the flip angle; the crusher gradients of the second/third inversion pulses have double/quadruple the area of the crusher gradients of the first inversion pulse, respectively, to prevent any stimulated echoes from constructively interfering with each other.
Since a spin-warp imaging trajectory was used for acquisition, all images are reconstructed with an inverse two-dimensional Discrete Fourier Transform.  Algorithm \ref{alg:multiMapAlg} and its accompanying descriptions detail the procedure for estimating the quantitative maps from these reconstructions.

A binary mask is created using using the average image (an example is shown in Fig. \ref{fig:mmBottles_2400_1200}) over all 22 images from the two segments using close and erosion morphological operations, similar to the methods presented in \cite{yang2015brain,pandian2008complex}.  Only those pixels indicated as having signal (white pixels in the mask) are processed in order to reduce the total computation time.

\setlength{\textfloatsep}{15pt}
\begin{algorithm}[ht]
  \protect
  \caption{multiMap's Quantitation Algorithm \label{alg:multiMapAlg}}

  \textbf{Inputs:} Image reconstructions ($I_1-I_{11}$ in figure \ref{fig:multiMapSeq}) from both segments, Intensity mask


  \textbf{Step}\hspace{8pt} \nl Estimate $B_1$ scaling factor with $I_8$ of both segments
    \label{alg:multiMapAlg_b1Mapping}

  \textbf{Step}\hspace{8pt} \nl Estimate $T_2$ with $I_9-I_{11}$ of the first segment
    \label{alg:multiMapAlg_t2}

  \textbf{Step}\hspace{8pt} \nl Estimate $\Delta B_0$, fat fraction, $T_2^\ast$ with $I_1-I_5$ of the first segment
    \label{alg:multiMapAlg_fatFrac}

  \textbf{Step}\hspace{8pt} \nl Estimate $T_1$ with $I_6-I_8$ of the first segment
    \label{alg:multiMapAlg_t1}

\end{algorithm}

\subsubsection*{\textbf{Step \ref{alg:multiMapAlg_b1Mapping}:} Estimate $B_1$}
The $I_8$ images of both segments are used to estimate $B_1$ scaling.  The value of $M_z\approx 0$ after saturation and increases during the recovery time.
The $30^\circ$ probing pulse that occurs during the saturation recovery time happens in both segments, so the $M_z$ at the end of the recovery time will be (approximately) equal.  Any remaining transverse magnetization is (approximately) eliminated by the spoiler near the end of the saturation recovery time.
Assuming ideal imaging pulses (meaning that the entire subject matter in each voxel experiences the same flip angle), the actual flip angle of the first imaging pulse can be calculated according to the double angle formula \cite{insko1993mapping} ($\alpha=\text{acos}\left( \sin(2\alpha) / 2\sin(\alpha) \right)$) and the $B_1$ scaling factor could be determined with $k=\alpha/(\pi/3)$.  However, since we are imaging a slice, the slice profile leads to errors in this estimate.  Instead, we create a lookup table (similar to the method presented in \cite{malik2011slice}): for a set of $B_1$ values, we calculate the slice profile of the $\alpha$ and $2\alpha$ pulses, we integrate across the slice and compute the ratio.  Given a ratio for a pixel of the reconstructed images, a reverse lookup into this table identifies the $B_1$ value that created this ratio.

Once the $B_1$ map is estimated, an average $B_1$ corrected image can be computed from all of the acquisitions.  This image can be used as an anatomical reference, and is presented in the results section as such.

\subsubsection*{\textbf{Step \ref{alg:multiMapAlg_t2}:} Estimate $T_2$}
Images $I_9-I_{11}$ of both segments along with the $B_1$ map of step \ref{alg:multiMapAlg_b1Mapping} are used to estimate $T_2$.  Nominally, the intensities of these images satisfy an exponential decay as described in section \ref{sec:background}.  However, this model neglects any errors in the flip angles of the inversion pulses \cite{majumdar1986errors}.  A more accurate estimate is obtained by modifying the signal model for the spin-echoes according to $\tilde{I}_{m,z} = M_{\xy}\left(t_{sat},z\right) \exp\left((t-t_{sat})/T_2\right) f_n(\theta(z))$ where $f_n$ is defined in table \ref{tbl:t2Modification} and $M_{\xy}(t_{sat},z)\in\mathbb{R}$ is the signal at location $z$ in the slice  \cite{majumdar1986errors}.  The rotation angle $\theta(z)$ is extracted from the composite matrix $R(z)$ \cite{slabaugh1999computing}.  The value $\tilde{I}_m$ is calculated from the slice profile by approximating the integral of \eqref{eq:sliceIntegral} with a Riemann sum as described in section \ref{sec:background}.
The value of $T_2$ is estimated for each pixel by solving the following optimization problem:
\begin{align*}
  \underset{ M_{\xy}(t_{sat}), \hspace{1pt} T_2 }{ \text{minimize} } & \hspace{2em} \left\| 
    \left(
      \left|I^{(1)}_9\right|,\left|I^{(1)}_{10}\right|,\left|I^{(1)}_{11}\right|
    \right) - \left(
      \tilde{I}_9,\tilde{I}_{10},\tilde{I}_{11}
    \right) \right\|_2 + \left\| \left(
      \left|I^{(2)}_9\right|,\left|I^{(2)}_{10}\right|,\left|I^{(2)}_{11}\right|
    \right) - \left(
      \tilde{I}_{9},\tilde{I}_{10},\tilde{I}_{11}
    \right) \right\|_2 \\
  \text{subject to} & \hspace{2em} T_2 > T_{2,\min} \geq 0 \text{ \hspace{1em} and \hspace{1em} } M_{\xy}(t_{sat}) > 0
\end{align*}
where $^{(i)}$ represents the data from the $i^{\text{th}}$ segment.
An initial estimate of the solution is found by linearly fitting the $\log$ of the data of the first segment to a line.  Interior point methods are used to solve this optimization problem.

\begin{table}
\centering
\begin{tabular}{|c|c|}
  \hline
  Spin Echo Index ($n$) & $f_n(\theta)$ \\
  \hline
  1 & $\frac{ (1-\cos(\theta)) }{ 2 }$ \\ 
  2 & $\frac{ (1-\cos(\theta))^2 }{ 4 }$ \\ 
  3 & $\frac{ (1-\cos(\theta))^3 }{ 8 } + \frac{ (1-\cos(\theta))(1+\cos(\theta))^2 }{ 8 } + \frac{ \cos(\theta)\sin(\theta)^2 }{ 2 }$ \\
  \hline
\end{tabular}
\captionsetup{justification=centering}
\caption{Spin Echo Signal Modification Function}
\label{tbl:t2Modification}
\end{table}

\subsubsection*{\textbf{Step \ref{alg:multiMapAlg_fatFrac}:} Estimate $\Delta B_0$, fat fraction, $T_2^\ast$}

Images $I_1$ and $I_3$ are used to determine an initial guess of the off-resonant frequency of each voxel: $\Delta \omega = \left( \text{angle}\left( I_1 \bar{I}_3 \right) \right) / \Delta t_{1,3}$, where $\bar{I}_3$ is the conjugate of $I_3$, and $\Delta t_{1,3}$ is the difference in acquisition times between the first and third images.
The angle between images $I_1$ and $I_3$ is higher than that of images $I_1$ and $I_2$, providing a higher angle-to-noise ratio.
It would be risky to use these images if there were potential for phase wrap in the calculation; a receiver bandwidth high enough to prevent phase wrapping between these images for samples of interest must be employed.  If this is not possible, images $I_1$ and $I_2$ could be used instead.

The signal model employed in this paper for images $I_1$ -- $I_5$, in accordance with \eqref{eq:mxyModel}, is
\begin{equation}
  I_m = \exp\left( i\Delta\omega_0 t_m \right) \left[
    \exp\left( -t_m / T_{2,W}^\ast \right) W + \exp\left( i\omega_{cs} t_m - t_m/T_{2,F}^\ast \right)  F
    \right].
  \label{eq:sigModel}
\end{equation}
In this expression, $t_m$ is the difference in time between the saturation pulse and the center of the acquisition for image $I_m$.  The parameters $W$, $F$, $T_{2,W}^\ast$, $T_{2,F}^\ast$, and $\Delta\omega_0$ are determined by solving the following optimization problem:
\begin{equation}
  \begin{aligned}
    \text{minimize} & \hspace{1em} \left\| \left( I_1, I_2, I_3, I_4, I_5 \right) - 
      \left( I^{(1)}_1, I^{(1)}_2, I^{(1)}_3, I^{(1)}_4, I^{(1)}_5 \right) \right\|_2 \\
    \text{subject to} & \hspace{1em} T^\ast_2 > T^\ast_{2,\min} \geq 0
                        \text{ \hspace{1em} and \hspace{1em} }
                        |\Delta \omega_0-\Delta\omega| < \Omega_0,
  \end{aligned}
  \label{eq:waterFatProb}
\end{equation}
where $\Omega_0>0$ is a bound on the observable off resonance frequency.  As the water or fat within the excited slice experiences the same exponential decay, we did not take the slice profile into account when estimating $T_2^\ast$.
In general, \eqref{eq:waterFatProb} is a non-convex optimization problem with several local minima.  However, if the $\Omega_0$ bound is small enough, then the optimal point is unique.

Note that if $\Delta\omega_0$, $T_{2,W}^\ast$, and $T_{2,F}^\ast$ were known, then $W$ and $F$ could be determined by solving the following linear system:
\begin{equation}
  \begin{aligned}
    \underbracket{ \begin{bmatrix}
      \exp\left(-t_1/T_{2,W}^\ast + i\Delta\omega_0 t_1 \right) & \exp\left(
        i(\Delta\omega_0 + \omega_{\text{cs}}) t_1 - t_1/T_{2,F}^\ast \right) \\
      \exp\left(-t_2/T_{2,W}^\ast + i\Delta\omega_0 t_2 \right) & \exp\left(
        i(\Delta\omega_0 + \omega_{\text{cs}}) t_2 - t_2/T_{2,F}^\ast \right) \\
      \exp\left(-t_3/T_{2,W}^\ast + i\Delta\omega_0 t_3 \right) & \exp\left(
        i(\Delta\omega_0 + \omega_{\text{cs}}) t_3 - t_3/T_{2,F}^\ast \right)
    \end{bmatrix} }_{\boldsymbol{A}} \underbrace{ \begin{bmatrix}
      W \\ F
    \end{bmatrix} }_{ \boldsymbol{\rho} } = \underbrace{ \begin{bmatrix}
      I_1 \\ I_2 \\ I_3
    \end{bmatrix} }_{ \boldsymbol{b} }.
  \end{aligned}
  \label{eq:waterFatLinSys}
\end{equation}
(This is similar to the linear system identified in \cite{reeder2004multicoil}.)
Estimates of $W$ and $F$ are determined with $\text{argmin}\|\boldsymbol{A}\boldsymbol{\rho}-\boldsymbol{b}\|_2$, which can be attained as follows: $\boldsymbol{\rho^\star} = \boldsymbol{A}^\dagger \boldsymbol{b}$, where $^\dagger$ indicates the pseudo inverse.  This can be accomplished in a numerically stable way using the QR decomposition or the Singular Value Decomposition \cite{trefethen1997numerical}.

To solve for all the optimization variables in \eqref{eq:waterFatProb} ($W$, $F$, $T_{2,W}^\ast$, $T_{2,F}^\ast$, and $\Delta\omega_0$), an exhaustive search is conducted over a discretized set of values for $T_{2,W}^\ast$, $T_{2,F}^\ast$, and $\Delta\omega_0$; for each triple of candidate values, the optimal $(W,F)$ is determined by solving \eqref{eq:waterFatLinSys}.  The set of parameters that achieve the lowest value of the objective function are the solution to the optimization problem.  Note that this is an embarrassingly parallelizable algorithm.
Once $W$ and $F$ are determined, the fat fraction is calculated as $f_f=|F|/(|W|+|F|)$.

From Bloch's equations, in a constant magnetic field, frequency is proportional to magnetic field (where the constant of proportionality is the gyromagnetic ratio) \cite{nishimura1996principles}: $\Delta \omega = \gamma \, \Delta B_0$.  Therefore, one can calculate the difference between the main magnetic field and the actual magnetic field using $\Delta B_0 = \Delta \omega / \gamma$.

\subsubsection*{\textbf{Step \ref{alg:multiMapAlg_t1}:} Estimate $T_1/M_0$ with $I_6$ -- $I_8$ of the first segment}

As described in section \ref{sec:background}, the $M_z$ component of the magnetization vector recovers according to $M_z(t,z) = M_z(t_0,z) + M_0\left( 1 - \exp\left( -(t-t_0) / T_1 \right), z \right)$, where $t_0$ is the starting time, $t>t_0$, $M_0$ is the proton density, and $z$ is the location in the slice.  After saturation, $M_z(t_0,z)\approx 0$ for all $z$; thus, $M_z(t,z)=M_0\left(1- \exp\left( -t / T_1 \right) \right)$, where $t$ is the time since the saturation pulse.

The flip angle of the probing pulses is $30^\circ$; this value was chosen to generate significant signal while limiting the amount that $M_z$ is altered to a small amount.  (Note that if one makes the small tip angle approximation \cite{nishimura1996principles} for the $30^\circ$ RF pulses then $M_0$ and $T_1$ can be determined by fitting the data to the exponential recovery model.  We have elected to use a more sophisticated model that does not require this approximation.)

The values of $T_1$ and $M_0$ are determined by solving the following optimization problem:
\begin{align}
  \underset{ T_1, \hspace{0.5pt} M_0 }{ \text{minimize} } & \hspace{2em} \left\| 
    \left(
      \left|I^{(1)}_4\right|, \left|I^{(1)}_6\right|, \left|I^{(1)}_8\right|
    \right) - \left( \hat{I}_4, \hat{I}_6, \hat{I}_8 \right) \right\|_2 \\
  \text{subject to} & \hspace{2em} T_1 > T_{1,\min} \geq 0 \text{ \hspace{1em} and \hspace{1em} } M_0 > 0.
  \label{eq:t1Obj}
\end{align}

When estimating $T_1$, we once again take the slice profile into account.  The signal model for each point in the slice of the reconstruction is
\begin{align*}
  M_{xy}(t_6,z) &= R_{30^\circ}(z) \left(
    0, 0, M_z(0) \left( 1 - e^{ -t_6 / T_1 } \right) + M_0\left( 1 - e^{ -t_4 / T_1 } \right)
    \right), \\
  M_{xy}(t_7,z) &= R_{30^\circ}(z) \left( 0, 0, M_{6,z}(z) \left( 1 - e^{ -(t_7-t_6) / T_1 } \right) + 
    M_0\left( 1 - e^{ -(t_7-t_6) / T_1 } \right) \right), \text{ and } \\
  M_{xy}(t_8,z) &= R_{60^\circ}(z) \left( 0, 0, M_{7,z}(z) \left( 1 - e^{ -(t_8-t_7) / T_1 } \right) + 
    M_0\left( 1 - e^{ -(t_8-t_7) / T_1 } \right) \right),
\end{align*}
where $R_{30^\circ}(z)$ and $R_{60^\circ}(z)$ are the composite rotation matrices for the nominally $30^\circ$ and $60^\circ$ RF pulses, respectively.  The times $t_6$, $t_7$, and $t_8$ are the times of the $6^{\text{th}}$, $7^{\text{th}}$, and $8^{\text{th}}$ acquisitions, respectively.   The value of $\hat{I}_6$ is determined by approximating the integral of \eqref{eq:sliceIntegral} with a Riemann sum (as discussed in section \ref{sec:background}); and similarly for $\hat{I}_7$ and $\hat{I}_8$.
The value of $M_z(0)\approx 0$; however, to account for any residual longitudinal magnetization after the saturation pulse, it is determined using the results of steps \ref{alg:multiMapAlg_b1Mapping} and \ref{alg:multiMapAlg_fatFrac}, according to $M_z(0) = ( |W| + |F| )/( k \, \tan\left( 90^\circ \right) )$, where $k$ is the $B_1$ scaling factor.

Interior point methods are used to solve \eqref{eq:t1Obj}.  In order to combat the tendency of the algorithm to result in a the location of a local minima, the optimization is conducted several times with different initial $T_1$ values.

Recall that when $x$ is small, $e^x\approx 1 + x$.  If the saturation recovery time is long enough that the signal has exited this linear regime, then both $T_1$ and $M_0$ can be estimated.  Note that due to the ambiguity in the scaling of the analog-to-digital converter of the system, we will only attain a value proportional to $M_0$.  If the saturation recovery time is not long enough to estimate these quantities individually, then one can assume the small tip angle approximation for the $30^\circ$ excitation pulses and fit the recovery to a line.  The slope of the line is $T_1/M_0$.  Alternatively, one can fit both $T_1$ and $M_0$ by minimizing \eqref{eq:t1Obj} and divide the resulting $T_1$ values by $M_0$.  It is this latter approach that we used to generate the results of this paper.

\subsection{Experiments}

All data was acquired on a $1.5$ T commercial scanner with linear gradient shimming.  We show results for three separate datasets: bottle phantoms, a knee, and a brain.  For the multiMap sequence, spoiler gradients imposed approximately four cycles across $4$ mm thick voxels.  The first/second/third pair of crusher gradients imposed $4$/$8$/$16$ cycles across the slice, respectively.  The excitation pulses were Hamming windowed sinc pulses with a time-bandwidth of $4$.  For the mask creation, the threshold was determined manually for each dataset independently.

\begin{figure}[ht]
  \centering{}
  \includegraphics[width=0.2\linewidth]{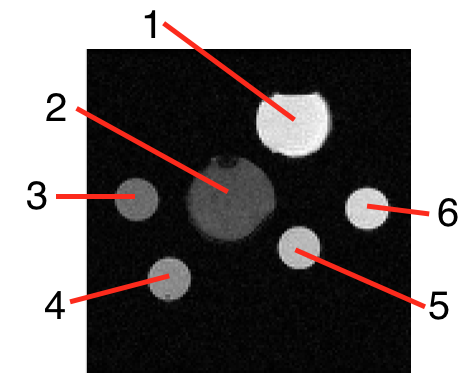}
  \caption{ An MR image of an axial slice of bottles used for testing multiMap.  Bottle 1 is filled with a copper sulfate solution.  Bottle 2 is filled with a manganese chloride solution.  Bottles 3, 4, 5, and 6 are filled with emulsified peanut oil in carrageenan gel with percentages of peanut oil by volume of $47\%, 29\%, 11\%$, and $0\%$, respectively. }
  \label{fig:bottleLabels}
\end{figure}

Images of bottles of size $128\times 128$ were collected with a field of view of $20\times 20$ cm using a $62.5$ kHz receiver bandwidth with a $62.5$ kHz receiver bandwidth and a $4.0$ mm slice thickness.  Data was collected of an axial slice of a set of $6$ bottles as shown in Fig. \ref{fig:bottleLabels}.  The bottles were filled with manganese chloride, copper sulfate, or emulsified peanut oil in carrageenan gel as specified in Fig. \ref{fig:bottleLabels}a. Peanut oil was used to simulate fat; it has a similar Larmor frequency.  The percentages of peanut oil in bottles $3$, $4$, $5$, and $6$ by volume (prior to curing) were $47\%$, $29\%$, $11\%$, and $0\%$, respectively.    The bottles were separated with MR compatible padding.  A single-channel quadrature birdcage head coil was used for both excitation and reception.

Images of a sagittal slice of a knee of size $128\times 128$ were collected with a $62.5$ kHz receiver bandwidth with a $62.5$ kHz receiver bandwidth and a $4.0$ mm slice thickness.  The field of view was $15\times 15$ cm.  The MR machine's body coil was used for excitation, and an extremity coil was used for reception.

Images of an axial slice of a brain of size $128\times 128$ were collected with a $62.5$ kHz receiver bandwidth with a $62.5$ kHz receiver bandwidth and a $4.0$ mm slice thickness.  The field of view was $20\times 20$ cm.  A single-channel quadrature birdcage head coil was used for both excitation and reception.

\section{Results}

\begin{figure}[ht]
  \centering{}
  \includegraphics[width=0.95\linewidth]{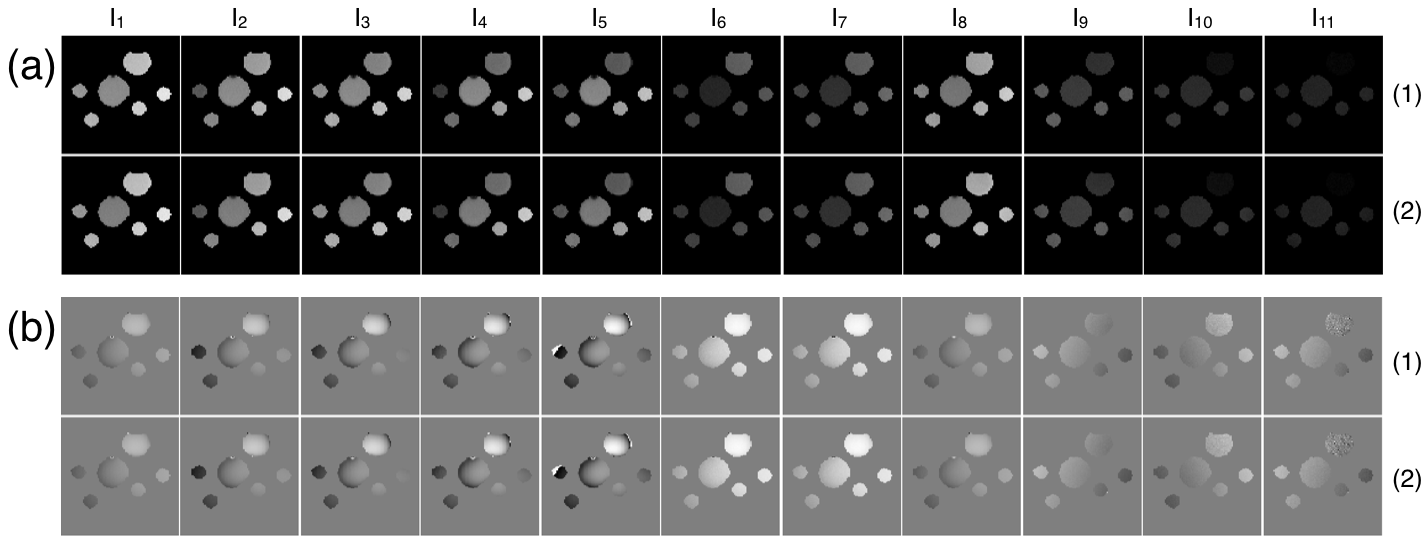}
  \protect\caption{The reconstructions of a slice through a set of bottles generated from the multiMap sequence; (a) are the magnitude images (in decibels) and (b) are the phase images.  The first and second rows of each sub-image are reconstructions from the first and second segments, respectively.}
  \label{fig:bottleRecons}
\end{figure}

Figure \ref{fig:bottleRecons} shows the $22$ axial slice images generated by the multiMap sequence. Figure \ref{fig:bottleRecons} (a) and (b) show the magnitude and phase images, respectively.  The top/bottom rows of each sub-image show the acquisitions of the first/second segments, respectively for a multiMap scan with a repetition time of $2400$ ms and a saturation recovery time of $1200$ ms.  The total scan time for this image was $10$ minutes and $34$ seconds.

\begin{figure}[ht]
  \centering{}
  \includegraphics[width=0.8\linewidth]{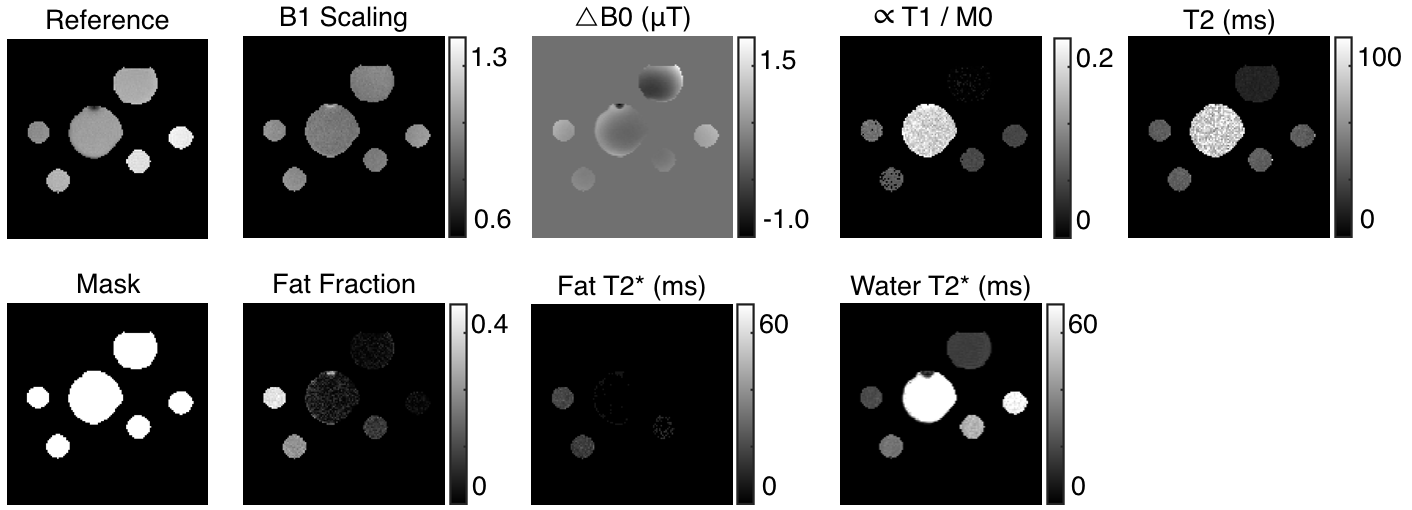}
  \protect\caption{Reference and quantitation maps generated from the data of the multiMap sequence with a repetition time of $2400$ ms and a saturation time of $1200$ ms.}
  \label{fig:mmBottles_2400_1200}
\end{figure}

Figure \ref{fig:mmBottles_2400_1200} shows the quantitative values estimated from the data of Fig. \ref{fig:bottleRecons} using the methods described in section \ref{sec:methods}.  The bright regions in the fat fraction map for bottles $3$, $4$, and $5$ are due to regions of the mask that exceed the actual data.  Note that the fat fraction of bottles $1$, $2$, and $6$ are all approximately $0$, as expected.

It is not expected that the fraction of signal due to fat equals the fat fraction volume (due to differences in the molar mass, differences in the number of hydrogen atoms per mole, and the chemical shielding of the fat molecule).  However, it is expected that the fraction of signal due to fat is proportional to the percentage of fat in the voxel.  We verify this in the fat fraction of the bottles estimated by comparing the ratios of different bottles as shown in Table \ref{tbl:fatFractionRatios}. The ratio of signal intensities is approximately equal to the ratio of fat fraction per volume.  This validates the fat fraction estimate attained by multiMap.

\begin{table}[ht]
  \centering{}
  \begin{tabular}{|c|c|c|c|}
    \hline
    Bottle Indices & Volume Ratio & Signal Ratio & Difference  \\ \hline
    $4$ : $3$ & $29 / 47$ = $0.6$ & $0.29 / 0.47$ = $0.6$ & $ <0.05$  \\ \hline
    $5$ : $4$ & $11 / 29$ = $0.4$ & $0.12 / 0.29$ = $0.4$ & $ <0.05$ \\ \hline
  \end{tabular}
  \caption{\label{tbl:fatFractionRatios} The table shows ratios of signal intensity and ratios of fat fractions in volumes.  The first column indicates the indices of the bottles analyzed.  The second column indicates the ratio of fat percentage in the bottles.  The third column is a ratio of average signal intensity of the bottles.  The fourth column shows the difference between the ratio of volumes and the ratio of signals.  For both pairs of bottles analyzed, the ratios are less that 0.05 different. }
\end{table}

\begin{figure}[ht]
  \centering{}
  \includegraphics[width=0.6\linewidth]{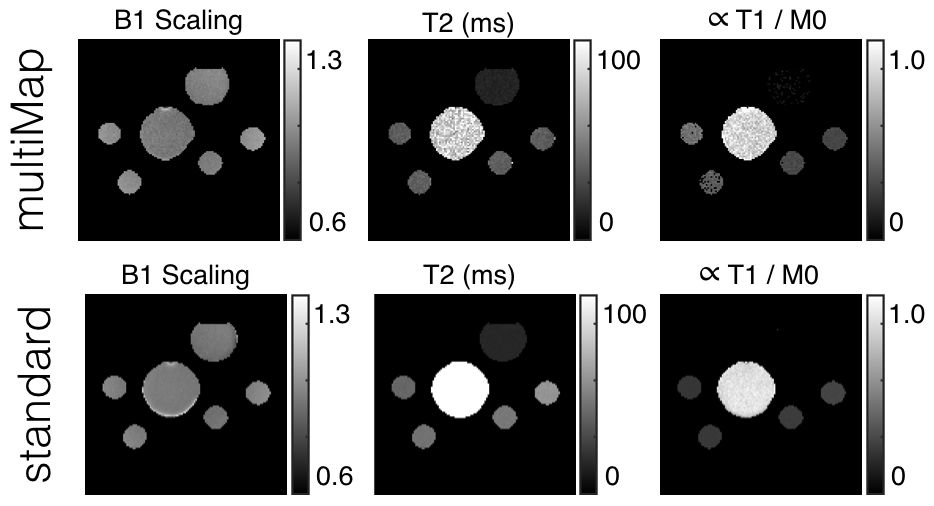}
  \caption{Standard Results.  a) Bottles $1$ is filled with Manganese Chloride dissolved in water.
  Bottle $2$ is filled with Copper Sulfate dissolved in water.
  Bottles $3$, $4$, $5$, and $6$ are made of emulsified peanut oil and carrageenan gel; the carrageenan gel is comprised of distilled water and 3\% carrageenan by weight. The percentages of peanut oil in bottles $1$, $2$, $3$, and $4$ by volume were $0\%$, $11\%$, $29\%$, and $47\%$, respectively.}
  \label{fig:standardComparison}
\end{figure}

We present the estimates of $B_1$, $\Delta B_0$, and $T_1/M0$ attained with multiMap to those of standard scanning methods in Fig. \ref{fig:standardComparison}.
Data was collected for each standard quantification individually.
A double angle mapping sequence was used to measure B1 \cite{cunningham2006saturated}.
An exponential decay was fit to a series of data collected from a spin-echo sequence with different echo times to estimate $T_2$.
A five-parameter model was fit to data from an inversion recovery sequence according to \cite{barral2010robust} in order to estimate $T_1$.
The figure shows good agreement for each quantity.
The estimate of $T_2$ of bottle $2$ is lower than the true value; this is due to the small length of time of the imaging pulse and the acquisitions $I_9$, $I_10$, and $I_{11}$.  In order to estimate longer $T_2$ values, one would need to spread out the acquisitions; this would come at a cost of reduced fidelity for smaller $T_2$ estimates.

For imaging the knee, the multiMap sequence used had a TR of $2400$ ms and a $t_{\text{sat}}$ of $1200$ ms.  The mask accurately isolates those pixels corresponding to tissue.  Under the assumption that $t_{\text{sat}}$ was long enough to accurately distinguish estimates of $T_1$ and $M_0$, we also presented $M_0$ imagery.  The fat and muscle show different $T_2$ values, as expected.
The $T_1/M_0$ quantity in the cortical bone is high due to the short $T_2$ leading to a near $0$ signal in this region in image $I_8$.
The fat fraction shows increased signal intensity in the fat and bone marrow, as expected.  

\begin{figure}[ht]
  \centering{}
  \includegraphics[width=0.8\linewidth]{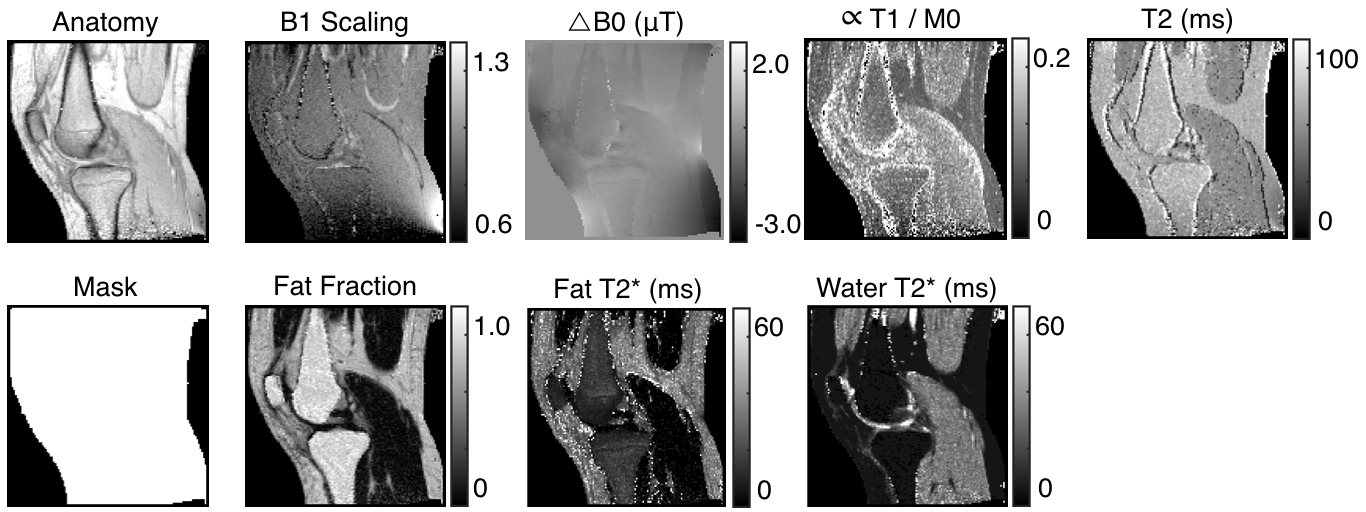}
  \protect\caption{Reference and quantitative maps of a sagittal slice of the knee determined with multiMap.  The windowing levels chosen for visualization reflects the sensitivity ranges based on the times of image acquisition.}
  \label{fig:mmKnee_2400_1200}
\end{figure}

For imaging the brain, the multiMap sequence used had a TR of $2400$ ms and a $t_{\text{sat}}$ of $1200$ ms.  The mask accurately isolates those pixels corresponding to tissue.  The $M0$ map accurately depicts similar proton density in brain parenchyma, and the ventricles and top of the brain stem show darker regions.  The $T_2$ map accurately depicts higher values in the ventricles.  The fat fraction map accurately identifies that the fat is largely isolated to subcutaneous tissue.  The ring of speckle in the fat fraction map (pointed to by the cyan arrow) results from the low signal intensity of bone.  The $T_2^\ast$ of water, mostly saturated, shows that this quantity is outside the estimable range by the multiMap sequence used do to the small time difference between the saturation pulse and image $I_5$.  To estimate larger $T_2^\ast$ values, one would need to either add additional images after the saturation pulse but before the first probing pulse or spread images $I_1-I_5$ to a longer times (this would come at a reduced fidelity of smaller $T_2^\ast$ estimates.

\begin{figure}[ht]
  \centering{}
  \includegraphics[width=0.9\linewidth]{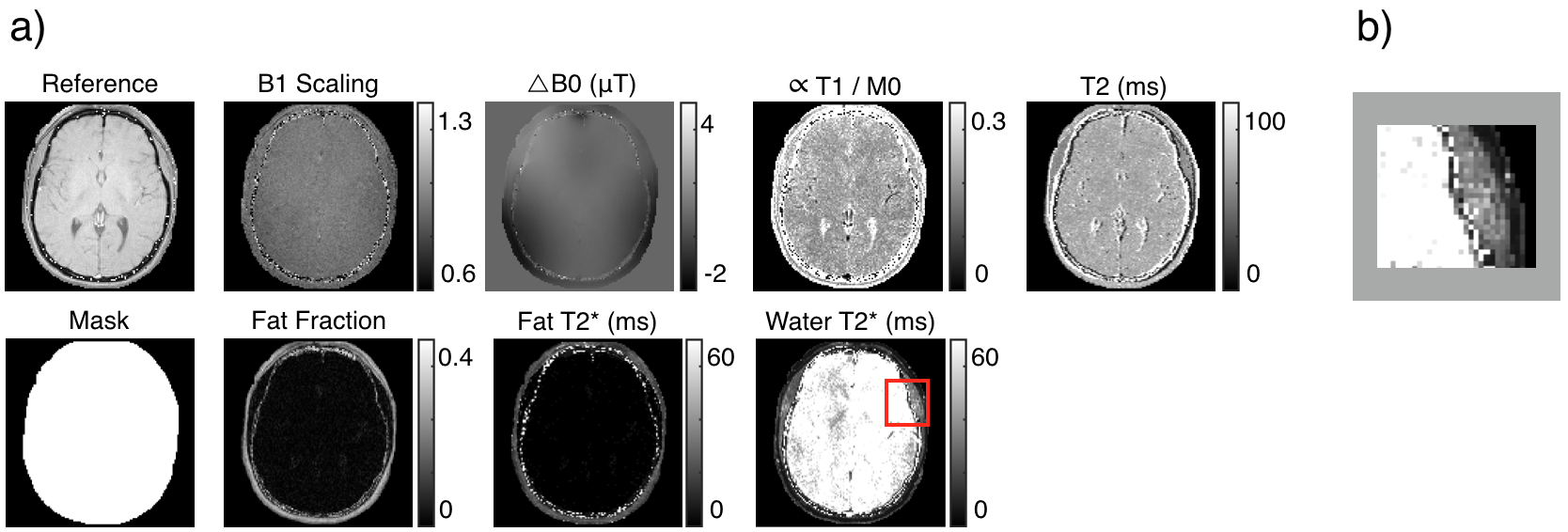}
  \protect\caption{ a) Reference and quantitative maps of an axial slice of the brain determined with multiMap.  b) A zoom-in of the red box of the water $T_2^\ast$ map.  There is a clear difference between the parenchyma of the brain and the subcutaneous fat.  The windowing levels chosen for visualization reflects the sensitivity ranges based on the times of image acquisition. }
  \label{fig:mmBrain_2400_1200}
\end{figure}

\section{Discussion}

The multiMap prescription combines several quantitative sequences into a single sequence.
It combines several standard and well-understood sequences that do not conflict with each other.
By isolating small sets of parameters into multiple estimation algorithms, it reduces cross-talk from the signals that could confound the estimation.
Moreover, since multiMap takes advantage of simple mathematical models that characterize the signal, there is no need for a large catalog of signals.

To increase the clinical utility, one could take advantage of scan reduction times offered by parallel imaging \cite{fessler2010model,pruessmann2006encoding} and compressed sensing \cite{lustig2007sparse,dwork2020calibrationless}.  Additionally, multiMap could be adapted to a multi-slice sequence; one could interleave preparation and readout to acquire additional volume in the same scan time.  We leave these extensions as possibilities of future work.

multiMap has the potential of performing a comprehensive analysis of the liver: elevated $T_2$ and $T_2^\ast$ are indicative of iron overload, elevated $T_1$ is indicative of Fibrosis or Cirrhosis \cite{yoon2016quantitative}, and elevated fat fraction is indicative of Steatosis.  In order to utilize this technique for this purpose, one would need to address the respiratory motion prior to image reconstruction \cite{pipe1999motion}.

One must be aware of the range of values one expects to image when using multiMap.  IF long $T_2$ values are required, for example, then acquisitions for images $I_{10}$ and $I_{11}$ must be far enough away from the imaging pulse to accurately quantify these estimates.  If there are large $T_2^\ast$ values expected, then one must similarly adjust the timings of acquisitions for images $I_1-I_5$.

In summary, we present multiMap, a single sequence for estimating several quantitative parameters.  We validate results on a set of bottle phantoms filled with a variety of substances.  We provide in-vivo results of a knee and a brain.

\section*{Acknowledgements}
The authors would like to thank Kirti Magudia for sharing her radiological expertise.

ND has been supported by the National Institute of Health's Grant Number P41 EB015891, the National Institute of Health's Grant Number T32EB009653 “Predoctoral Training in Biomedical Imaging at Stanford University”, the National Institute of Health's Grant Number NIH T32 HL007846, the Sloan Fellowship, the Rose Hills Foundation Graduate Engineering Fellowship, the Electrical Engineering Department New Projects Graduate Fellowship, and The Oswald G. Villard Jr. Engineering Fellowship.

JP has been supported by the National Institute of Health's Grant Number P41 EB015891.


\section*{Conflicts of Interest}
JP is on the advisory board of Heart Vista Inc.


\end{document}